\documentclass[%
 reprint,
 amsmath,amssymb,
 aps,
pra,
]{revtex4-2}
\usepackage{multirow}
\usepackage{algorithm}
\usepackage{makecell}
\usepackage{algpseudocode}
\usepackage{float}
\usepackage[justification=raggedright,singlelinecheck=false]{caption}
\usepackage{array}    % p{}, m{}形式を利用
\usepackage{physics}
\usepackage{braket}
\usepackage{txfonts}
\usepackage{url}
\usepackage{color}
\usepackage{bm}
\usepackage[pdftex]{graphicx}
\usepackage{url}

\bibliography{physRevA_submissionNotes}

\begin{document}

\preprint{APS/123-QED}

\title{Transfer Learning for Deep-Unfolded Combinatorial Optimization Solver {with} Quantum Annealer}
\author{Ryo Hagiwara, Shunta Arai, and Satoshi Takabe}
\affiliation{{Institute of Science Tokyo}, Ookayama, Tokyo 152-8550, Japan}
\date{\today}

\begin{abstract} 
Quantum annealing (QA) has attracted {research} interest as a sampler and combinatorial optimization problem (COP) solver.
{A recently proposed} sampling-based solver for QA significantly {reduces} the required number of qubits, {being capable of large COPs}.
{In relation to this}, a trainable sampling-based COP solver {has been proposed that} optimizes its internal parameters from a dataset by using a deep learning technique called deep unfolding.
Although learning {the} internal parameters accelerates the convergence speed, the sampler in the trainable solver is restricted to {using} a classical sampler {owing} to the training cost.
In this study, to utilize QA in the trainable solver, we propose classical-quantum transfer learning, where parameters are trained classically, and the trained parameters are used in the solver with QA.
{The results of numerical} experiments demonstrate that the trainable quantum COP solver using classical-quantum transfer learning improves convergence speed and execution time {over the original solver}.
\end{abstract}
\maketitle

\section{\label{sec:level1}INTRODUCTION}

A combinatorial optimization problem (COP) {entails determining} the minimum (or maximum) solution to an objective function with discrete variables under given {constraints}. 
Typical examples {of COPs} include the traveling salesman and knapsack problems. 
COPs are related to the P-NP problem, and {in} many cases, finding {an} optimal solution in polynomial time is considered difficult~\cite{PNP}.

{Two {well-known} optimization approaches are simulated annealing (SA)~\cite{SA} and quantum annealing (QA)~\cite{QA}}.
{In particular}, QA has attracted significant {research} attention {owing to its ability to sometimes} outperform classical optimization algorithms, such as SA, in terms of convergence speed and {solution quality}~\cite{QAvsSA}.
QA has been implemented in specialized devices, such as the quantum annealer~\cite{DwaveSys}.
{However, the quantum annealer faces limitations such as {a} relatively small number of available qubits and fixed qubit couplings {within} the device.}
Mapping problem constraints onto the hardware's limited qubit connectivity {of hardware} often restricts the size and complexity of the instances that can be practically addressed~\cite{MinerEmbedding1, MinerEmbedding2, MinerEmbedding3, MinerEmbedding4}.

{To mitigate these limitations,} a new sampling-based COP solver using gradient methods has been proposed~\cite{Ohzeki}, which we refer to as the Ohzeki method {in this paper}. 
{In the Ohzeki method, a constrained problem is re-formulated as a sampling problem with auxiliary variables introduced by the Hubbard-Stratonovich transformation~\cite{Hubbard,Stratonovich}.
Then, the method {then iteratively} executes two {processes}: a sampling {process} for the original variables and a gradient descent {process} for the auxiliary variables.
{This} enables {the} quantum annealer to {handle bigger problems by avoiding the} costly mapping from constraints to a device. 
In addition, any sampler is applicable to the Ohzeki method, whereas the method was originally proposed for a quantum annealer.
A drawback of the method is tuning step sizes in the gradient descent {process}. 
A heuristic {selection} of step sizes often leads to performance degradation.}

{{In our previous study}~\cite{duom}, {we tuned} the step sizes of the Ohzeki method in a \textit{data-driven} {manner}. 
{{This} concept {in machine learning, known as} deep unfolding (DU)~\cite{DU,DU2}}, uses techniques such as backpropagation to optimize the internal parameters of differentiable iterative algorithms.}
{Because} DU is based on iterative algorithms, it provides an easy interpretation of the training parameters.
Additionally, {because} {only a few internal parameters must be learned, the learning costs are reduced.}
DU has been shown to improve convergence speed and approximation performance compared to the base algorithms~\cite{ChebyshevSteps}.
{Consequently,} DU has been employed to develop novel optimization-based {trainable algorithms} for signal processing, including areas such as wireless communications~\cite{DUCS}, compressed sensing~\cite{TISTA}, and image processing~\cite{DUIM}.
{For example, {DU has been employed to handle} COPs {arising} in wireless communication tasks such as digital signal detection, and DU has been employed for these detection algorithms, achieving remarkable performance improvements~\cite{DUCS,THS}.}
In addition, some Deep-unfolded classical algorithms inspired by quantum computation have {also} been proposed~\cite{TSB,DULQA}. 
{Overall,} DU is a powerful model-based learning approach {that improves} existing algorithms.

{In our previous work~\cite{duom}, we combined the Ohzeki method with DU to tune the step sizes, which {affected the} convergence speed and optimization performance.
{Specifically,} automatic differentiation and backpropagation {cannot be used} when the learning step sizes. 
To circumvent this, we {formulated} a new learning strategy {wherein} a sampler estimates gradient information {is approximated} through a sampling part. 
We verified that the trainable solver, {namely} the deep-unfolded Ohzeki method (DUOM) was {sufficiently trained} using {this} strategy. 
In addition, DUOM accelerates the convergence {of} the original Ohzeki method with a constant step size, indicating the effectiveness of DU for a sampling-based algorithm.
{However, a} limitation of {this approach} is the use of the classical Metropolis-Hastings (MH) algorithm~\cite*{Metropolis,Hastings}, {rather than} QA {{for} sampling}.
{This} is inevitable because DUOM {executes} many times in the training process, and {its} training cost of DUOM with a quantum annealer {is} impractically high.}

{The aim of this paper is to propose and examine a learning strategy for  DUOM with a quantum annealer to circumvent the costly training process. 
Specifically, we propose a \textit{classical-quantum transfer learning} where the trainable parameters are {preliminarily learned by DUOM with a classical sampler, and subsequently used to execute DUOM with a quantum annealer.} 
We {investigated} the effect of a classical sampler {on} the training process by comparing the MH algorithm with simulated quantum annealing (SQA)~\cite{SQA, SQA2}, which emulates QA on a classical computer.} 
We {also compared} the performance of the classical-quantum transfer learning {with that of} conventional parameter tuning by grid search. 
{In addition, the execution time of DUOM with a quantum annealer {was} compared to that {of} a classical sampler to examine the quantum acceleration.}

{
The {remainder} of this paper is {organized} as follows. 
Section~\ref*{SEC_OM} {introduces the Ohzeki method for constrained binary quadratic optimization problems}. 
Section~\ref*{SEC_DUOH} provides {an overview of the DU and DUOM proposed in our previous {study}}. 
Section~\ref*{SEC_CQTL} describes classical-quantum transfer learning.
Section~\ref*{SEC_EXPERIMENTS} {{presents} the results of numerical experiments {conducted}} the performance of classical-quantum transfer learning.
Finally, Section~\ref*{SEC_CONCLUSION} concludes the paper.}

\section{\label{SEC_OM}Ohzeki Method}
To solve COPs using a quantum annealer, {an optimization problem is} converted into a {quadratic unconstrained binary optimization (QUBO)} format~\cite{QUBO}.
{The penalty method is typically used when} expressing a COP with linear constraints in QUBO form~\cite{QUBO}.

Consider the following QUBO with linear constraints
\begin{equation}
  \begin{aligned}
  \min_{\bm{x} \in \{0,1\}^N} \quad  & f_0(\bm{x}) \\
  \textrm{s.t.} \quad & f_k(\bm{x}) = C_k \quad (k=1, \dots ,m),
  \end{aligned}\label{eq_COP}
\end{equation}
where $f_0(\bm{x})$ is the objective function representing the cost or energy to be minimized, 
{and $f_k(\bm{x})=C_k$ is the $k$-th constraints in which $f_k$ is a linear function.}
This formulation {can be used to express} practical COPs such as the traveling salesman problem and the graph partitioning problem. 
To convert COP~(\ref{eq_COP}) into {the} QUBO form, the penalty method introduces the following {loss function} by adding a penalty term to the objective function $f_0(\bm{x})$:
\begin{equation}
  L_\lambda(\bm{x}) = f_0(\bm{x}) + \lambda \sum_{k=1}^m(f_k(\bm{x}) - C_k)^{2},
  \label{loss}
\end{equation}
where $\lambda > 0$ is a hyperparameter controlling the {{penalty} strength}. 
{If a solution} satisfies the constraints of COP~(\ref{eq_COP}), the penalty term $\lambda \sum_{k=1}^m(f_k(\bm{x}) - C_k)^2$ in the loss function becomes zero.

{Although} this penalty method effectively converts COP~(\ref{eq_COP}) into the QUBO form,
it also increases the number of qubits required for QA, {placing an upper bound on the size of problems solvable by QA}. 
{That is,} the number of quadratic terms in the QUBO increases, requiring additional qubits for minor embedding~\cite{MinerEmbedding1, MinerEmbedding2, MinerEmbedding3, MinerEmbedding4} {{to} map} logical variables onto the physical qubits of {a} quantum annealer.

To address this {issue}, Ohzeki proposed a solver~\cite{Ohzeki} that uses the Hubbard-Stratonovich transformation~\cite{Hubbard, Stratonovich} to linearize the penalty terms.
{Let us consider solving {the} constrained COP~(\ref{eq_COP}) by sampling from the {Boltzmann} distribution corresponding to Eq. (\ref{loss}),  defined by}
\begin{equation} \label{OhzekiBoltmannDist}
  Q(\bm{x}) = \frac{1}{Z} \exp(-\beta f_0(\bm{x}) - \beta \lambda \sum_{k=1}^{m} (f_k(\bm{x}) - C_k)^2),
  \nonumber
\end{equation}
{where} $\beta$ is the inverse temperature, {which should be sufficiently large for optimization}, and $Z$ is the partition function. 
{{Alternatively, the} Ohzeki method samples from a new Boltzmann distribution with auxiliary variables $\bm{v}$ introduced by the Hubbard-Stratonovich transformation~\cite{Ohzeki}.  
The distribution is given by}
\begin{equation} \label{OhzekiBoltmannDist}
  Q(\bm{x}; \bm{v}^{(t)}) = \frac{1}{Z(\bm{v}^{(t)})} \exp(-\beta f_0(\bm{x}) + \beta \sum_{k=1}^{m} v_k^{(t)} f_k(\bm{x})).
\end{equation}
{A sampler is used to estimate the expectations $\{\braket{f_k(\bm{x})}_{Q(\bm{x}; \bm{v}^{(t)})}\}_{k=1}^{m}$ of functions $\{f_k(\bm{x})\}_{k=1}^{m}$ over the distribution (\ref{OhzekiBoltmannDist}).
Then, the auxiliary variables $\bm{v}$ are updated to satisfy $\braket{f_k(\bm{x})}_{Q(\bm{x}; \bm{v}^{(t)})}=C_k$ for each $k$. 
{This is achieved by gradient descent with the following update rule for each $k${:}}}
\begin{equation}
  v_k^{(t+1)} = v_k^{(t)} + \eta_t (C_k - \braket{f_k(\bm{x})}_{Q(\bm{x}; \bm{v}^{(t)})}) \quad (t = 0, 1, \dots), \label{eq_OH} 
\end{equation}
{where} $\eta_t$ {the step size at iteration $t$ for the gradient descent method.}

 \begin{figure}[!t]
\begin{algorithm}[H]
    \captionsetup{justification=raggedright, singlelinecheck=false}
  \caption{Ohzeki method}\label{alg_oh} 
  \begin{algorithmic}[1] 
  \State \textbf{Input:} $\beta$,$\lambda$,$f_0$, $\{f_k, C_k\}$, max iteration: $T$
  \State \textbf{Initialize:} $\bm{v}^{(0)} \in \mathbb{R}^m$
  \For{ $t = 0, 1, \dots ,T-1$}
  \State estimate $\{\braket{f_k(\bm{x})}_{Q(\bm{x}; \bm{v}^{(t)})}\}$ by sampling from
  \State \quad $Q(\bm{x}; \bm{v}^{(t)}) \propto \exp(-\beta f_0(\bm{x}) + \beta \sum_{k=1}^{m} v_{k}^{(t)} f_k(\bm{x}))$
  \State update $\bm{v}^{(t+1)}$ by
  \State \quad $v_k^{(t+1)} = v_k^{(t)} + \eta_{t} ( C_k - \braket{f_k(\bm{x})}_{Q(\bm{x}; \bm{v}^{(t)})})$
  \EndFor\label{euclidendwhile}
  \State \textbf{return}  $\arg\min L(\bm{x}; \lambda)$ by sampling from $Q(\bm{x}; \bm{v}^{(T)})$
  \end{algorithmic}
\end{algorithm}
 \end{figure}

Algorithm~\ref{alg_oh} presents the pseudocode of the Ohzeki method.
{Using this method}, the number of required qubits is reduced, allowing {a quantum annealer to handle larger constrained COPs}.

\section{\label{SEC_DUOH}Deep-Unfolded Ohzeki Method}

{In this section, we briefly summarize the concept of DU. 
{We then} introduce the deep-unfolded Ohzeki method, {wherein} the internal parameters are learned in a data-riven manner.}

\subsection{\label{SEC_DUOH_DU}Deep Unfolding}

DU is a deep learning technique that optimizes internal parameters in iterative algorithms by embedding trainable variables and applying backpropagation. 
{Here, we illustrate for gradient descent as an example. For interested readers, please refer to {previous} reviews~\cite{Monga, Boyd}.} 

\begin{figure}[t]
  \centering
  \includegraphics[trim=0 0 0 0,width=0.5 \textwidth]{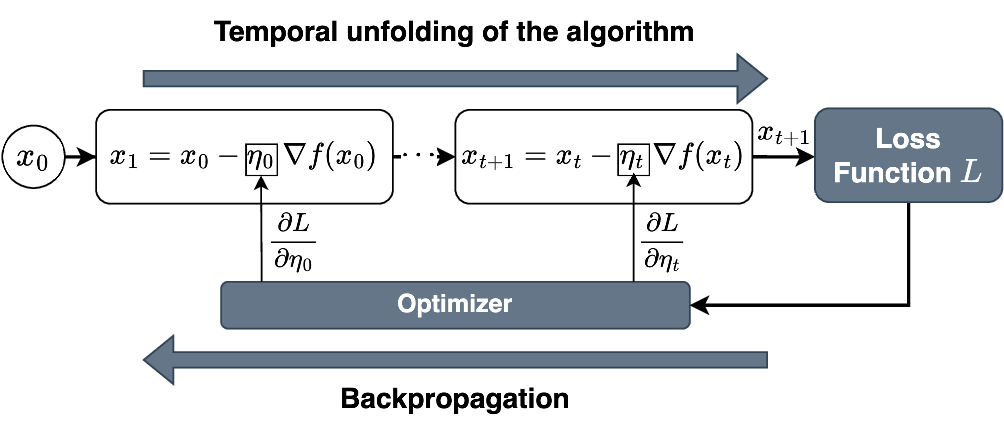}
  \caption{
    A schematic diagram of {deep-unfolded gradient descent (upper) and its training process (lower)}.
    }
  \label{deep_unfolding}
\end{figure}

In gradient descent, the update rule for variable $\bm{x}$ at iteration {$t =0,1, \dots, T-1$} is given by
\begin{equation}
\bm{x}_{t+1} = \bm{x}_t - \eta_t \nabla f(\bm{x}),
\nonumber
\end{equation}
where $T$ is the total number of iterations, $\eta_t$ is a step size, {and $f$ is a random objective function, such as a quadratic function with random coefficients, to be minimized.}
{The selection of} an appropriate step size is crucial; {an excessively} large step size may {result in failure to converge, whereas an overly} small step size may result in slow convergence or {cause the algorithm to fall into local minima}.
In DU, the iterative process is temporally unfolded, embedding trainable parameters to optimize performance. 
{In this case,} trainable step sizes ${\{\eta_t\}}^{T-1}_{t=0}$ {are introduced as {shown} in Fig.~\ref*{deep_unfolding}.}

Assume that a dataset $\{\bm x_0^{(d)},f^{(d)}(\bm x),\bm x_*^{(d)}\}_{d=1}^D$ is available, where $D$ represents the number of problem instances in the dataset.
Each instance contains an initial point $\bm x_0^{(d)}$, an objective function $f^{(d)}(\bm x)$, and corresponding optimal solution $\bm x_*^{(d)}$.
{{This} setting corresponds to supervised learning}.
To learn $\{\eta_t\}^{T-1}_{t=0}$ suitable for the dataset, a loss function $L$ is introduced.
For instance, the mean squared error (MSE) loss function is defined {as}
\begin{equation}
L= \frac{1}{D}\sum_{d=1}^D \|\bm x_T^{(d)}-\bm x_\ast^{(d)}\|_2^2,
\end{equation}
where $\bm x_T^{(d)}$ is an output of {a} gradient descent {with} initial point $f^{(d)}(\bm x)$. In other words, the loss function for DU minimizes the distance between the output obtained {using} a given trainable parameters and the optimal solutions.
Then, using backpropagation, the gradients $\partial L / \partial \eta_t$ are calculated to {find optimal step sizes that minimize the loss function.
{This} learning process {illustrated by the} lower arrow in Fig.~\ref*{deep_unfolding}.
As a result, {the} learned gradient descent accelerates convergence and improves} approximation performance compared {with that of the} original gradient descent~\cite{Chebyshev}.

\subsection{\label{SEC_DUOH_DUOH}Deep-Unfolded Ohzeki Method}
As mentioned {previously}, the Ohzeki method is an efficient COP solver {that uses} gradient descent.
However, {because} the Ohzeki method applies gradient descent (\ref{eq_OH}) to a non-convex {objective function}, its convergence speed and performance depend on the step size, {which {is selected} heuristically}.
To address this {issue}, the authors {have proposed} the deep-unfolded Ohzeki Method (DUOM) in~\cite{duom}, which learns {step sizes based {upon} DU}.

%The details of DUOM are explained below.
The learnable parameters in the DUOM are the step sizes $\{\eta_t\}_{t=0}^{T-1}$ {in Eq.~(\ref*{eq_OH})}.
The DUOM adjusts the step sizes to minimize the loss function~(\ref*{loss}), aiming to satisfy the constraints and minimize the objective function.
{Unlike the supervised learning {described} in the {previous} subsection, DUOM adopts \textit{unsupervised learning} {which} does not require optimal solutions {for} a dataset.}
{Unsupervised learning is suitable owing to the challenge of estimating optimal solutions for constrained COPs in advance.}
{Then,} the step sizes are updated using {an optimizer such as Adam~\cite{Adam}}, with gradients obtained by backpropagation.
The partial derivative of the loss function $L_\lambda$ with respect to $\eta_t$ is given by
\begin{align}
\frac{\partial L_\lambda}{\partial \eta_{t}} &= \sum_{k=1}^{m}\frac{\partial L}{\partial v_k^{(T)}}  \left(\prod_{u=t+1}^{T-1} \frac{\partial  v_k^{(u+1)}}{\partial{v_k^{(u)}}}\right)
\frac{\partial v_k^{(t+1)}}{\partial \eta_t} , \label{eq_d1}\\
\frac{\partial v_k^{(u+1)}}{\partial v_k^{(u)}} &= 1 - \eta_{u} \frac{\partial \braket{f_k(\bm{x})}_{Q(\bm{v^{(u)}})}}{\partial v_k^{(u)}}. \label{eq_d2}
\end{align}
{Because} {the expectation $\braket{f_k(\bm{x})}_{Q(\bm{x}; \bm{v}^{(t)})}$ is estimated {via}} non-differentiable sampling, automatic differentiation {of} deep-learning packages such as PyTorch~\cite{pytorch} {is inapplicable to Eq.~(\ref{eq_d2})}.
Although {this gradient can be approximated by numerical differentiation}, {this} is computationally expensive and inefficient {{as} it requires additional sampling processes}.
{{As an alternative}, in~\cite{duom}, the authors proposed the sampling-based gradient estimation via the variance. Using Eq.~(\ref{OhzekiBoltmannDist}), we have}
\begin{equation}
\frac{\partial \braket{f_k(\bm{x})}_{Q(\bm{x}; \bm{v}^{(t)})}}{\partial v_k} = \beta \left (\braket{f_k^2(\bm x)}_{Q(\bm{x}; \bm{v}^{(t)})} - \braket{f_k(\bm x)}_{Q(\bm{x}; \bm{v}^{(t)})}^2 \right ).
\label{eq_ex}
\end{equation}
{This allows {for} gradient estimation in the forward process. {That is,} the gradient is efficiently calculated {using} the variance, which {approximated when} sampling the expectation over $Q(\bm{x}; \bm{v}^{(t)})$}.

In~\cite{duom}, {the} numerical results showed that DUOM exhibited excellent performance compared with {that of} the conventional Ohezeki method. 
However, {DUOM could only be deployed with a classical sampler} because {the} training process required a {large} number of sampling processes.
{In the numerical experiments on image reconstruction, {the} sampling process was executed {approximately one} million times during training.}
Although a quantum annealer is applicable to DUOM in principle, {the use of} QA in the training process is impractically {expensive}. 

\section{\label{SEC_CQTL}classical-quantum transfer learning}
{The aim of this {study} is to examine DUOM {using a quantum annealer without} without QA {during} the training process. 
The key observation is that any sampler is applicable to DUOM.} 

{We propose classical-quantum transfer learning as a new learning strategy. 
In the scheme, in general, an unfolded algorithm is trained using a classical (approximate) algorithm, and the learned parameters are then transferred to that with a quantum algorithm.} 
{{Unlike conventional} transfer learning, {where a model trained on a given} dataset is applied to another type of dataset, classical-quantum transfer learning uses classical computation in the training process and quantum computation in execution. 
The difference in algorithms between training and execution {is considered to incur} performance degradation, {as the} learned parameters may be unstable {in} quantum algorithms. 
However, if the difference is sufficiently small, the performance degradation can be reduced. 
{This reflects} the importance of {selecting} a classical algorithm {for} the training process.}

\begin{figure}[t]
  \centering 
\includegraphics[trim=0 0 0 0,width=0.5 \textwidth]{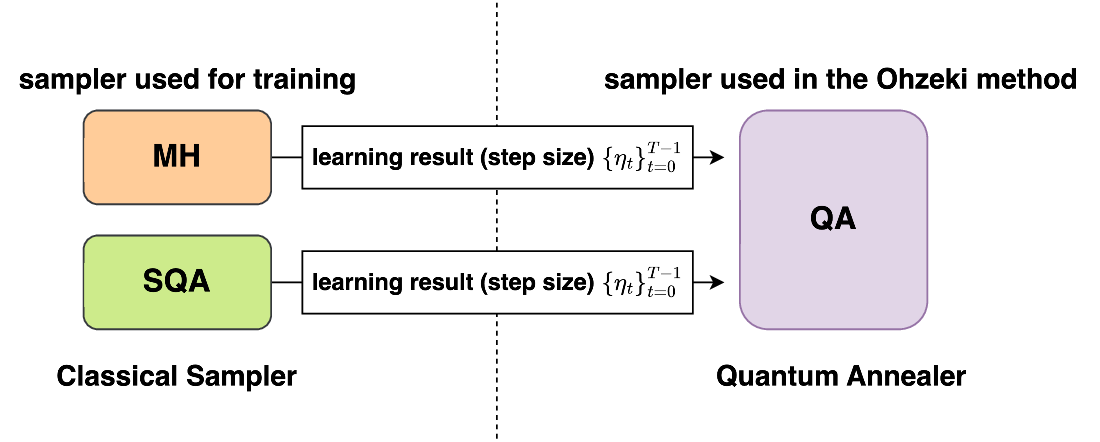}
  \caption{Overview of classical-quantum transfer learning, {where learned parameters using a classical sampler (left) are transferred to DUOM with a quantum annealer in execution (right).}    
}
  \label{transfer_lerning}
\end{figure}

{In our case, classical-quantum transfer learning {entains} that 
the trained step sizes of DUOM with a classical sampler are transferred to DUOM with a quantum annealer. 
In particular, we examined the MH algorithm and SQA as classical {samplers} used for training. 
Figure~\ref*{transfer_lerning} {presents} an overview of classical-quantum transfer learning.
This approach significantly reduces {costs} by eliminating quantum computation during {the} training process.}

It {should be} emphasized that the proposed classical-quantum transfer learning differs from other techniques developed in quantum computing. 
For example, the quantum approximate optimization algorithm (QAOA)~\cite{QAOA} {handles COPs by tuning the} internal parameters related to {the} quantum gates.
{{When the QAOA parameters are tuned}, the objective function is evaluated by executing a quantum circuit multiple times.
{Classical Bayesian or gradient-based optimization}, {with gradients evaluated using the parameter shift method}~\cite{ParamShiftMethod}, {is} then used for parameter updates. 
The feedback loop between quantum circuit execution and classical optimization in QAOA differs from {that in} classical-quantum transfer learning, {wherein} the training process {does not involve quantum computation}. 
In other words, the QAOA parameter tuning process is not classical-quantum transfer learning because the parameter update depends on outputs of quantum computation.}
On the other hand, classical-quantum transfer learning does not involve  quantum computation during the training process.
Instead, the training is performed {entirely} on a classical computer, and the learned parameters are then applied to quantum computations to improve performance.

{Another related method {involves} optimizing the initial states of quantum circuits {using} classical tensor networks~\cite{Tensor}.
Although {this} approach {appears} similar to classical-quantum transfer learning, {it} focuses only on the initial state.
In high-dimensional quantum computation~\cite{MPS},the limitations of the expressive power and high computational costs of tensor networks {limit the tuning process to only these initial states}.
{Conversely,} in the classical-quantum transfer learning {configuration of} DUOM, any internal parameters {can be learned} by replacing QA with a classical and computationally cheap sampler.
To the best of our knowledge, the proposed classical-quantum transfer learning {algorithm} is the first attempt {at} internal parameter tuning to reduce computational costs without {using} quantum computation in the training process.}

\section{\label{SEC_EXPERIMENTS}Experiments}

{In this section, we {present} numerical experiments {conducted} on an image reconstruction problem to verify {the performance of} classical-quantum transfer learning for DUOM. 
First,} we focus on comparing MH {with SQA, which approximates QA, as a classical sampler for training}. 
{Next, we examined the Trotter number in SQA, which {controls} the accuracy {of the QA approximation}. 
Finally, we compared the performance of DUOM with QA to that {of} a classical sampler during execution {to analyze the effects of QA.}} 

\subsection{\label{sec:level2}Problem and parameter settings}
We focused on a two-dimensional binary image reconstruction {using} linear measurements.
Let $\bm{x} \in \{0,1\}^N$ be a vector representing a binary image of {size} $\sqrt{N} \times \sqrt{N}$. 
{This} problem aims to recover $\bm{x}$ from the noiseless linear observations $\bm{y} = \bm{A}\bm{x}$, where $\bm{A} = (a_{kl})$ is an $M \times N$ random matrix ($M < N$) whose elements are independently drawn from a standard normal distribution.
Assuming that {the} non-zero pixels in the original image are neighboring {pixels}, the problem is defined {as} 
\begin{equation}
  \min_{\bm{x} \in \{0,1\}^N} \,\,  - \sum_{\braket{i,j}} x_i x_j \,\, 
     \textrm{s.t.} \quad  y_k = \sum_{l=1}^{N}a_{kl}x_l \,\, (k=1, \dots ,M),
\end{equation}
where $\langle i,j \rangle$ denotes adjacent pixel pairs in the image.  
This formulation corresponds to Eq.~(\ref{eq_COP}) with $f_0(x) = -\sum_{\langle i,j \rangle} x_i x_j$, $f_k(x) = \sum_{l=1}^N a_{kl} x_l$, and $C_k = y_k$ for $k = 1, \ldots, M$.

\begin{figure}[t]
    \centering
    \includegraphics[trim=50 250 0 250,width=0.5 \textwidth]{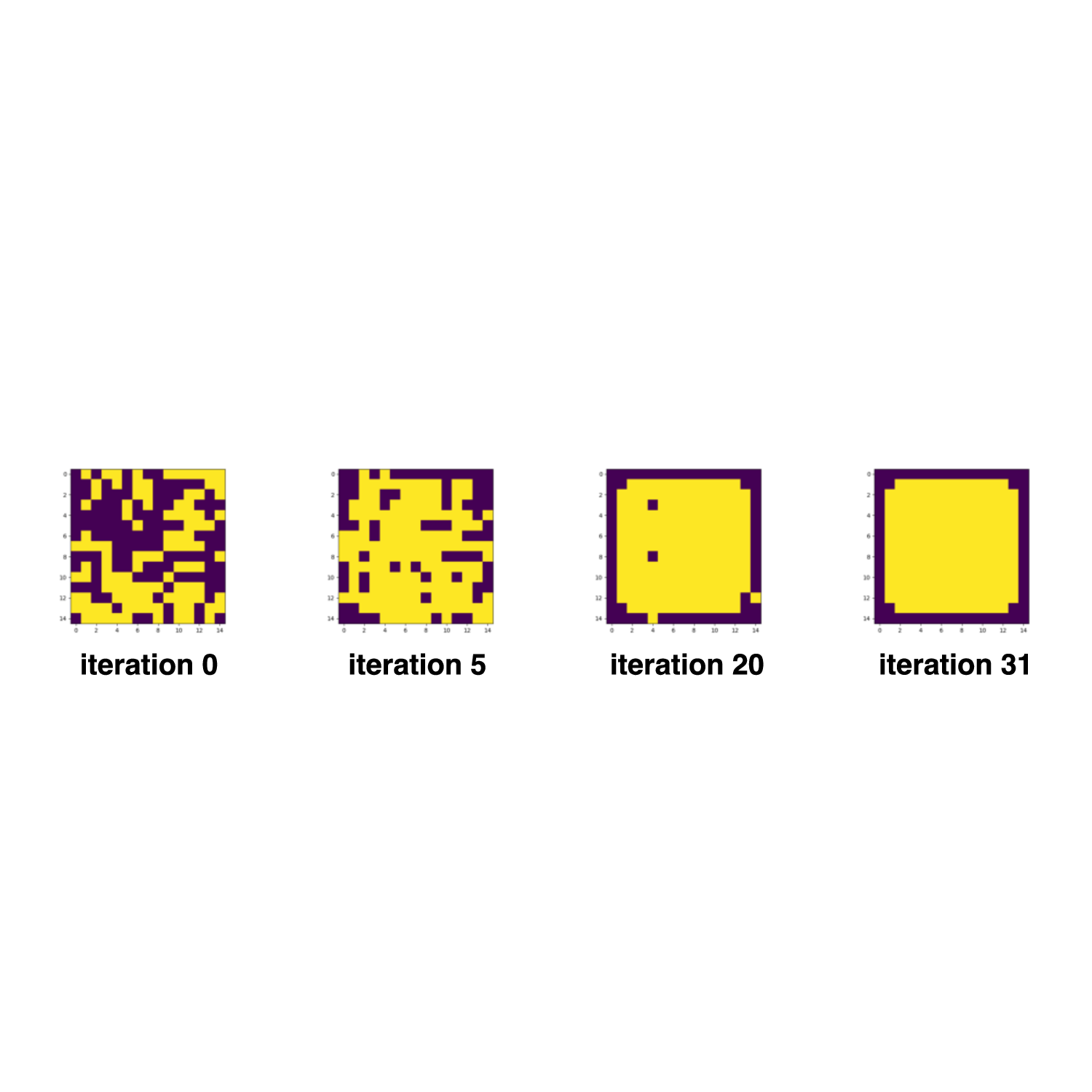}
    \caption{
        Examples of reconstructed images using the Ohzeki method with $\eta = 1.0 \times 10^{-2}$. 
        In each image of size $15 \times 15$, yellow and purple represent $x_i = 1$ and $x_i = 0$, respectively. 
        From left to right, the images correspond to iterations $0$, $5$, $20$, and $31$, 
        where the rightmost image is equivalent to the original one.
    }
    \label{img_reco}
  \end{figure}
In the experiment, we set $N=15^2=225$ and $M=135$. 
{The ratio $M/N = 0.6$ is lower than the statistical-mechanical  threshold $M/N=0.633$~\cite{TANAKA}. 
It suggests that the COP is typically hard to solve in the thermodynamic limit ($N \rightarrow \infty$).}

{To implement} unsupervised learning {for} DUOM, we generated datasets consisting of random matrices $\bm{A}$ and corresponding observations $\bm{y} = \bm{A} \bm{x}^\ast$, with the original image $\bm{x}^\ast$ fixed throughout the experiments, as illustrated in Fig.~\ref*{img_reco}.
In each parameter update, we used $20$ mini-batches of size $4$.
The parameters ${\{\eta_t\}}_{t=0}^{T-1}$ were optimized using the Adam optimizer~\cite{Adam} to minimize the loss function in Eq.~(\ref{loss}) of   $\lambda = 1$.
The initial learning rate was set to $5.0 \times 10^{-2}$ and was decayed by a factor of $0.8$ at each iteration to enhance convergence.
Incremental learning techniques~\cite{Incremental1} were applied to prevent the gradient vanishing.
The total number of iterations was set to $T = 30$, with initial parameter values of $\eta_t = 1.0 \times 10^{-2}$ for all $t$.

{{To implement} SQA,} {we used} SQASampler from OpenJij~\cite{OpneJij}, with the Trotter number set to the default value of $4$ layers, unless otherwise noted.
For {comparison}, the DUOM using MH {was} {trained and executed} under the same conditions {defined} in the previous {study}~\cite{duom}.
We used the Advantage\_system $6.4$ of the D-Wave machine {as a quantum annealer}. 

\subsection{\label{SEC_NUMERICAL_RESULTS}Numerical results}
\subsubsection{\label{sec:level3}Performance of DUOM with quantum annealer}
First, we evaluated the performance of {classical-quantum} transfer learning.
{We compared three methods that uses QA in {their} execution: DUOM trained using SQA (SQA-QA), DUOM trained using MH (MH-QA), and the original Ohzeki method {with a} constant step size was optimized by a grid search.}

\begin{figure}[t]
  \centering
  \includegraphics[trim=0 0 0 0,width=0.5 \textwidth]{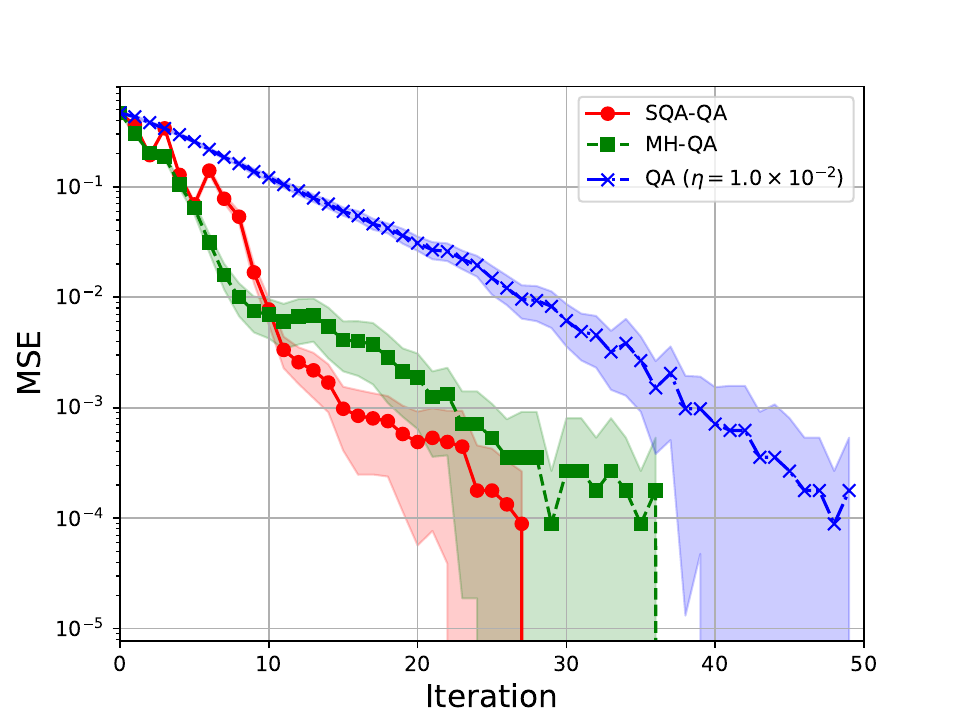}
  \caption{
    {MSE performance of DUOM by classical-quantum transfer learning as a function of the number of iterations. SQA-QA and MH-QA represent DUOM trained using SQA and MH, respectively.}}
  \label{DUOM_ex1_img}
\end{figure}

Figure~\ref{DUOM_ex1_img} illustrates {performance results of these methods in terms of MSE}. 
At each iteration, we computed the best MSE obtained from the sampler's {outputs} and then averaged these values over $50$ random instances. 
The same instances were used consistently throughout {the} numerical experiments. 
The error bars {represent} $95$ percent confidence intervals of the data.
{These results} indicated that the performance of SQA-QA, which transferred the training results from SQA to QA, {achieved the highest performance}, with MSE reaching zero within $30$ iterations.
In comparison, using a fixed step size without training did not reduce the MSE to zero within $50$ iterations, {indicating} a significant performance improvement {by classical-quantum transfer learning}. 
Moreover, {SQA-QA even outperformed} MH-QA, which transferred the training results from MH to QA.
This is {possibly because SQA can approximated QA more accurately than MH,} making {the} transfer learning to QA more effective.
To {examine} this, we next compared the performance based on the Trotter number of SQA.

\subsubsection{\label{sec:level3} Trotter number dependency in SQA}
{SQA approximates QA by using multiple Trotter layers, {as} introduced by the Suzuki-Trotter decomposition~\cite{Trotter, Suzuki}.}
{The Trotter number $\tau$ is a hyperparameter {that controls the accuracy of the approximation}.
{{Given a limit of} $\tau \rightarrow \infty$}, SQA reproduces the original quantum system described by the Schrödinger equation, as required by the Suzuki-Trotter formula~\cite{Trotter, Suzuki}.
However, it is important to note that the dynamics of SQA do not fully replicate the physical dynamics of QA, {which are} governed by the Schrödinger equation~\cite{YukiBando}.
}

\begin{figure}[t]
  \centering
  \includegraphics[trim=0 0 0 0,width=0.5 \textwidth]{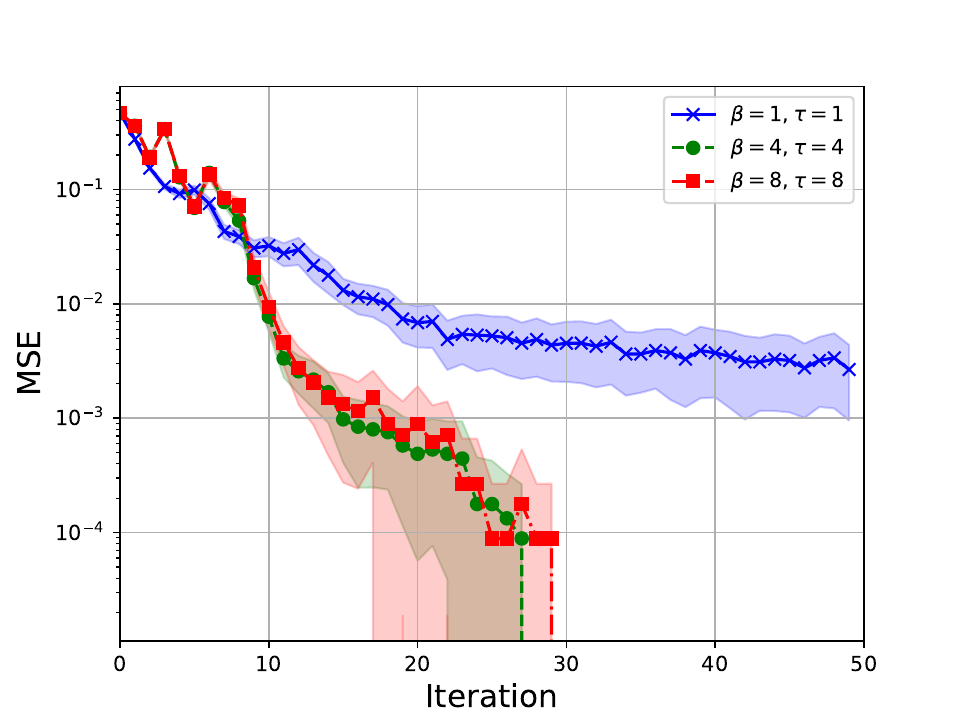}
  \caption{{Trotter number $\tau$ dependency of MSE performance for SQA-QA when $\beta=\tau$.}}
  \label{duom_Trotter}
\end{figure}

{We conducted numerical experiments to examine the effect of the Trotter number on classical-quantum transfer learning.}
Figure~\ref*{duom_Trotter} presents the numerical results when $\beta=\tau$.
The MSE and the error bars are plotted in the same {manner}, as {shown} in Fig.~\ref*{DUOM_ex1_img}.
{Note that SQA with $\beta = \tau =1$ is equivalent to MH with the inverse temperature $\beta=1$.}
{In this case,} the MSE {failed to} reach zero within $50$ iterations, indicating poor performance.
{In contrast}, when the Trotter number was increased to $\tau = 4$, {DUOM {achieved} optimal solutions for all instances} within $30$ iterations.
Increasing the Trotter number to $\tau=8$ yielded {nearly equivalent performance to that} with $\tau=4$, {{whereas} optimal solutions were found by DUOM with fewer iterations for {certain} instances.}
This suggests that {a large Trotter number in {the} SQA in the training process leads to better performance in classical-quantum transfer learning}.

\subsubsection{\label{sec:level3} Comparison of quantum DUOM with classical DUOM}
{We now compare DUOM performance with different samplers used during execution} 
We considered three scenarios: 
DUOM trained using SQA {and} executed on QA (SQA-QA), DUOM trained {and} executed using SQA (SQA-SQA), and DUOM trained {and} executed using MH (MH-MH).
{Note that SQA-SQA and MH-MH represent conventional classical DUOM {configurations} without transfer learning, whereas SQA-QA involved classical-quantum transfer learning.}

\begin{figure}[t]
  \centering
  \includegraphics[trim=0 0 0 0,width=0.5 \textwidth]{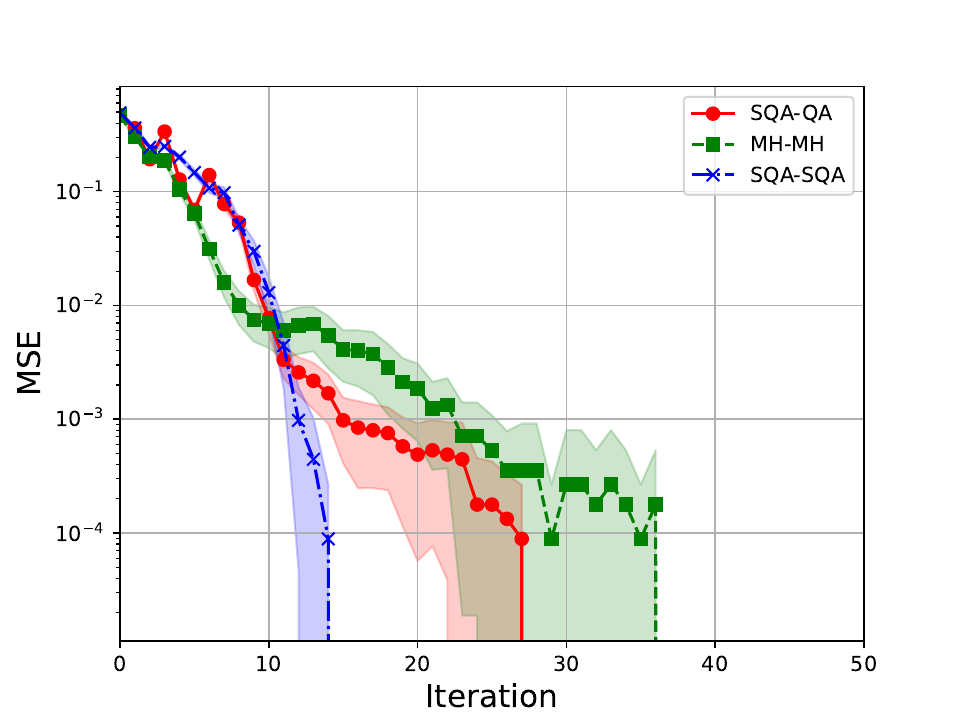}
  \caption{MSE performance of DUOM by classical-quantum transfer learning (SQA-QA) with classical DUOM with MH (MH-MH) and SQA (SQA-SQA).
    }
  \label{DUOM_ex2_img}
\end{figure}

Figure~\ref{DUOM_ex2_img} {presents the results of this experiment in terms of MSE.}
The MSE and the error bars are plotted in the same {manner, shown} as in Fig.~\ref*{DUOM_ex1_img}.
{Here, the} SQA-SQA method {exhibited} the best performance, with all instances converging to {the} optimal solutions within $15$ iterations. 
In comparison, SQA-QA converged more slowly, requiring more iterations to {obtain} optimal solutions. 
The MH-MH exhibited the slowest convergence among them.
The superior performance of SQA-SQA can be attributed to the consistency between the sampler used during training and execution. 
{Because} both {the} training and execution processes used SQA, the learned parameters were optimal for the SQA sampler. 
{In contrast, SQA-QA involves transfer learning, where the algorithm used during training process differs from that used during execution.}
This discrepancy may lead to suboptimal performance {compared to the SQA-SQA configuration.}

{We also examined execution time as another important metric for evaluating COP solvers.}
{Specifically}, we {compared} the three methods in terms of the execution time. 
Table \ref*{execution_time_table} summarizes the average execution {times required to solve} a single instance. 
{This} table includes the average sampling time, gradient descent time, total execution time without {an} API, API response time for D-Wave, and the total execution time.
\begin{figure}[t]
    \centering
    \begin{table}[H]
        \centering
        \begin{tabular}{|c|c|c|c|c|c|}
        \hline
        & Sampling       & GD            & Total w/o API & API          & Total        \\
        \hline
        \multirow{2}{*}{SQA-QA} & $181$ ms & $108$ $\mu$s & $181$ ms & $12.2$ s & $12.2$ s \\
        % \cline{2-6}
                                & ($53.1$ ms) & ($32.9$ ns)  & ($53.1$ ms) &  ($4.14$ s) & ($4.14$ s) \\
        \hline
        \multirow{2}{*}{SQA-SQA} & $34.8$ s & $180$ $\mu$s & $34.9$ s &  --    & $34.9$ s \\
        % \cline{2-6}
                                &($4.44$ s)&  ($43.3$ ns)   &($4.44$ s)&     --     &($4.44$ s)\\
        \hline
        \end{tabular}
        \caption{{Average (upper) and standard deviation (lower) of execution time for SQA-QA and SQA-SQA.}}
        \label{execution_time_table}
    \end{table}
\end{figure}
{Despite requiring more iterations, SQA-QA was significantly faster in terms of the total execution time {than} SQA-SQA.
When the API response time {was} ignored, the execution time of SQA-QA {was approximately} $190$ times faster than {that of} SQA-SQA.
In particular, the sampling time of SQA-QA {was} negligible compared to that of SQA-SQA, whose execution time {was} dominated by the SQA sampler.
Even if the API response time is considered, the total execution time for SQA-QA {was} $12.2$ seconds, {whereas that for SQA-SQA was} $34.9$ seconds.}
Although SQA-SQA converges in fewer iterations, the rapid sampling capability of the quantum annealer in SQA-QA {significantly reduces the} total execution time. 
This speed-up highlights the practical advantage of using a quantum annealer for execution, even when {employing} transfer learning. 
{These} results demonstrate that classical-quantum transfer learning can leverage the speed of quantum hardware to {accelerate} optimization, despite potential performance differences in convergence behavior.

\section{\label{SEC_CONCLUSION}CONCLUSION}
In this {study}, we proposed a novel {trainable COP} solver that {integrates} QA with DU {using} classical-quantum transfer learning. 
This approach effectively {mitigates} the high training {costs} associated with quantum computation by leveraging classical computation, {thereby} making quantum computation more accessible and efficient for deep-unfolded quantum algorithms.
{The results of} numerical experiments on the image reconstruction problem demonstrated the practical viability and effectiveness of the proposed method.
Specifically, the transfer learning approach, {wherein} step sizes are learned using a classical sampler and transferred to a quantum annealer, showed a significant improvement in convergence speed and accuracy.
Classical-quantum transfer learning is a classical-quantum hybrid method {designed for} trainable quantum algorithms.

The results highlighted the {effect} of using classical samplers {such as} MH and SQA, for training.
The comparisons revealed that SQA, which better approximates QA, {achieved} superior performance when its training parameters were transferred to a quantum annealer.
This indicates that {selecting} an appropriate classical sampler is crucial for effective classical-quantum transfer learning.
{In addition}, the influence of the Trotter number in SQA on the accuracy of QA approximations was examined, showing that larger Trotter numbers improved the performance of classical-quantum transfer learning.
This opens up new avenues for exploring different classical approximations to enhance learning efficiency.

Furthermore, the comparisons between classical and quantum DUOM {demonstrated} that {although} SQA-SQA {requires} fewer iterations to converge compared to SQA-QA, the latter {achieved} a clear advantage in terms of execution time, emphasizing the practical utility and speed of quantum execution {following training}.
These findings {underscore} the feasibility of deploying quantum-enhanced solvers trained on classical systems for real-world optimization tasks.

Future research directions include the following.
First, it is crucial to verify the effectiveness of classical-quantum transfer learning and quantum DUOM for more practical problems, {including the} traveling salesman and graph partitioning problems.
The application of this approach to COPs with inequality constraints, such as the knapsack problem~\cite{Ohzeki2}, is another crucial area for further investigation.

\begin{acknowledgments}
    {This study was partially supported by JSPS KAKENHI with Grants No. 22H00514 and No. 22K17964.}
\end{acknowledgments}

\end{document}